\documentclass{jfm}

\usepackage{graphicx}
\usepackage{epstopdf, epsfig}
\usepackage{float}
\usepackage{caption}
\usepackage{subcaption}
\usepackage{amsmath}
\usepackage{amsfonts}
\usepackage{breqn}
\usepackage{nicefrac}

\newcommand{\abs}[1] {|#1|}

\shorttitle{2D Convection-Diffusion in Multipolar Flows}
\shortauthor{E. Boulais and T. Gervais}

\title{2D Convection-Diffusion in Multipolar Flows}

\author{Etienne Boulais\aff{1}
 \and Thomas Gervais\aff{1}}

\affiliation{\aff{1}Department Engineering Physics, Polytechnique Montreal,
2500 Chemin de Polytechnique, Montreal, QC H3T 1J4}

\begin{document}

\maketitle

\begin{abstract}
We present a complete analysis of the problem of convection-diffusion in low $\Rey$, 2-dimensional flows with distributions of singularities, such as those found in open-space microfluidics and in groundwater flows. Using Boussinesq transformations and solving the problem in streamline coordinates, we obtain concentration profiles in flows with complex arrangements of sources and sinks for both high and low $\Pen$. These yield the complete analytical concentration profile at every point in applications that previously relied on material surface tracking, local lump models or numerical analysis such as microfluidic probes, groundwater heat pumps, or diffusive flows in porous media. Using conformal transforms, we generate families of symmetrical solutions from simple ones, and provide a general methodology that can be used to analyze any arrangement of source and sinks. The solutions obtained that contain the explicit dependence on the various parameters of the problems, such as $\Pen$, the spacing of the apertures and their relative injection and aspiration rates. In particular, we show that the high $\Pen$ models can model problems with $\Pen$ as low as 1 with a maximum error comitted of under $10\%$, and that this error decreases approximately as $\Pen^{-1.5}$.
\end{abstract}

\section{Introduction}
Many 2-dimensional or quasi 2-dimensional flow problems can be approximated as arrangement of point sources and sinks in an unconfined 2D space. Transport in such planar multipolar flows is important in many areas of fluid mechanics. Multipolar flows appear in impinging jet problems, for example in industrial cooling processes \citep{webb1995single} or in the study of diffusion flames \citep{spalding1961theory}. Transport in such flows has also been analyzed in the context of hydrocarbon recovery \citep{koplik1994tracer} \citep{kurowski1994anomalous} \citep{chen1998miscible} and in the exploitation of aquifers in hydrogeology \citep{luo2004fluid} \citep{luo2007breakthrough}. Distributions of multipoles have been used as approximations for flow past rigid obstacles in the context of convection trough porous media \citep{eames1999longitudinal}, for the modelling of fluid interface in porous media \citep{de1960singularity}, for approximating transport of suspended particles in quasi-2D colloids \citep{cui2004anomalous} or for the modelling of droplet suspensions in microfluidic channels \citep{beatus2006phonons}. In biology, multipolar flows can be found in the irrigation of the choriocapillaris \citep{zouache2016form}. In engineering, many devices include flow from thin pipes injected to a thin space between two plates. These include hydrostatic thrust bearings \citep{jackson1965pressure}, radial diffusers \citep{woolard1957theoretical} and injection molding processes \citep{kamal1972injection}. Among multipolar flows, quadrupole flows are of special interest as the flow within them is purely extentional. They have been used to study the stretching of oil droplets \citep{taylor1934formation} and as exact solutions of the Navier-Stokes equation for the study of vortex stretching \citep{burgers1948mathematical} \citep{bazant2005exact}. 

More recently, in the context of microfluidics, there has been a renewed interest in the study of passive transport in multipolar laminar flows. Arrangements of sources and sinks have been used to generate hydrodynamically confined flows with microfluidic probes \citep{juncker2005multipurpose} \citep{autebert2014hierarchical} or to generate tunable concentration gradients in microfluidic quadrupoles \citep{qasaimeh2011microfluidic}. Similar devices have also been used in a non-confined operation mode to generate omnidirectional concentration gradients \citep{nakajima2016microfluidic}. Outside of open-space devices, multipolar flow has been used to model intricate flow patterns in microchannel intersections \citep{lee2007microfluidic} \citep{shenoy2016stokes}, and can be used to model planar concentration gradient generators \citep{atencia2009microfluidic}. Microfluidic multipoles have also been used as a building block to create reconfigurable concentration patterns on open surfaces \citep{goyette2019microfluidic}.

In light of these applications, there is a need for a deeper understanding of 2D transport processes in multipolar flows. Considerable work has already been done on the subject: Transit times have been found for particles in symmetrical dipole flows both without diffusion effects \citep{luo2004fluid}, as well as with diffusion for some geometries \citep{koplik1994tracer}, scaling laws have been found for various physical parameters in microfluidic dipoles using linearized equations near the flow’s stagnation point \citep{safavieh2015two}, analysis of the gradient at the center of microfluidic quadrupoles have yielded precise expressions for diffusion length \citep{qasaimeh2011microfluidic} and thorough numerical simulations have been done for several multipole geometries \citep{christ2011design} \citep{boulais2018hele}. More recently, Zouache and Eames published an investigation of flow and passive transport in the turning region of multipolar flows as well as in triangular subcells of tesselated flows, such as are found in the human choriocapillaris \citep{zouache2019flow}. However, despite these advances, there is still a need for a complete framework for the analysis of general 2D multipolar transport problems. Analytic approaches usually either neglect diffusion entirely and focus on material surface tracking \citep{da1960pattern} \citep{grove1970fluid}, or find diffusion lengths for very specific regions of space \citep{qasaimeh2011microfluidic} \citep{safavieh2015two}. Neither of these methods allows for the analysis of complex 2-dimensional flow profiles and concentration gradients. On the numerical side, calculations are made laborious by the highly multiscale aspect of multipolar flow problems \citep{boulais2018hele}, which puts strict constraints on the meshing of the geometry and makes calculations very resource intensive. Recently, we presented a theoretical model for the complete 2D concentration profile in microfluidic multipoles at high $\Pen$ as well as a method for generating more elaborate solutions using conformal transforms \citep{goyette2019microfluidic}. This model was restricted to high Peclet flows with hydrodynamic confinement (meaning problems in which the net aspiration rate is larger than the net injection rate), as is found in microfluidics applications. In this paper, we present a complete analysis of the problem of 2D convection-diffusion in planar multipolar flows at low $\Rey$. We then give solutions to the problem for both high and low $\Pen$ regimes, valid for problems both with and without hydrodynamic confinement. Combined with conformal transforms, 2D multipolar problems can serve as a building block to understand a great variety of flow configurations. We find that in practice, the ”high” $\Pen$ approximation yields precise results for values of $\Pen$ as low as 1.

\section{Theory}

\subsection{Hele-Shaw flow}

Starting from the incompressible Navier-Stokes equation, we can demonstrate that, for a flow of sufficiently small Reynolds number confined in a thin space between two parallel plates, the velocity field is \citep{batchelor2000introduction}

\begin{equation}
	\begin{aligned}
	u &= - \frac{1}{2 \mu} \frac{\partial p}{\partial x} z \left(d - z\right) \\
	v &= - \frac{1}{2 \mu} \frac{\partial p}{\partial y} z \left(d - z\right) \\
	w &= 0
	\end{aligned}
\end{equation}

Which is the product of a potential flow in the xy direction and a parabolic profile in the z direction. Averaging over the z direction, we get streamlines that are exactly analogous to field lines in a 2D electric field \citep{hele1898flow}. The 2D flow field can thus be written as

\begin{equation}
	\pmb{u} = \nabla \phi
\end{equation}

With the potential $\phi$ proportional to the pressure field, and obeying

\begin{equation} \label{laplace_eq}
	\nabla^2 \phi = 0
\end{equation}

We can use Gauss’ theorem to obtain the adimensional potential generated by a point source or sink

\begin{equation}
	\phi_i = Q_i \text{log} \left( \abs{\pmb{x} - \pmb{x_i}} \right)
\end{equation}

Where $Q_i$ is the adimensional flow rate at the aperture and $\pmb{x_i}$ is its position. Since the governing equations are linear, the flow generated by a set of point sources and sinks is simply given by the sum of the contribution of each source and sink.
Transport of a passive tracer such as heat or a dilute molecule in the flow is described by the adimensional convection-diffusion equation. At steady-state, supposing no flux of $c$ through either the top or bottom wall, this equation is

\begin{equation} \label{convection_diffusion_eq}
	\nabla^2 c - \Pen \ \nabla \phi \cdot \nabla c = 0
\end{equation}

Solving this equation allows us to obtain the 2D concentration profile in multipolar flows. Equation \ref{convection_diffusion_eq} neglects the 3D aspect of the flow and is thus inappropriate for capturing effects along the z-axis such as diffusion very near the turning region, effects of prescribed concentration on the top or bottom plate \citep{zouache2019flow} or diffusion broadening due to the butterfly effect \citep{ismagilov2000experimental}. However, the concentration profile described by the 2-dimensional convection-diffusion equation remains very appropriate in high Peclet number flows in microfluidic applications, where the region of diffusive broadening due to no-slip condition can be made very small, and in Darcy flow applications where the no-slip condition on the top and bottom don’t actually apply. In cases where more precise information is needed about the 3D concentration profile in a specific region of the flow, the 2D solution could be used as a baseline to which an analysis similar to \citep{ismagilov2000experimental}, analogous to the Leveque solution \citep{leveque1928lois} is added as a perturbation.

\subsection{Complex potential}

We can rewrite our position vectors as complex numbers $z = x + i y$ and redefine our spatial derivative operators using

\begin{equation}
	\nabla = \frac{\partial}{\partial x} + i \frac{\partial}{\partial y}
\end{equation}

This enables us to express the complex potential $\phi$ as a function of a single variable $z$. The contribution of point sources becomes simply

\begin{equation}
	\Phi_i = Q_i \text{log} \left( z - z_i \right)
\end{equation}

The gradient of the real part of the complex potential gives the velocity field, while the imaginary part gives the streamlines of the flow. The main advantage of expressing $\phi$ as a function of a complex variable is that the system consisting of equations \ref{laplace_eq} and \ref{convection_diffusion_eq} is conformally invariant \citep{boussinesq1902pouvoir} \citep{bazant2004conformal}. This means that we can use conformal transforms to map simple known solutions to more complex geometries. This will be the main approach we use when solving transport problems in multipolar flows.

\section{Multipolar flows}

We are interested in convection-diffusion in multipolar flows, that is flows generated by discrete sets of point-like injection and aspiration apertures. In practice, these can be fabricated by having small circular channels deliver or aspirate fluid from the top plate of the Hele-Shaw cell. The difference in flow profile between a finite circular aperture and a point-like source is negligible outside of the radius of the aperture \citep{boulais2018hele}.

\subsection{Finite-size dipoles}\label{section_dipole}

\begin{figure}

	\centering
	\begin{subfigure}{0.45\textwidth}
		\includegraphics[width = \textwidth]{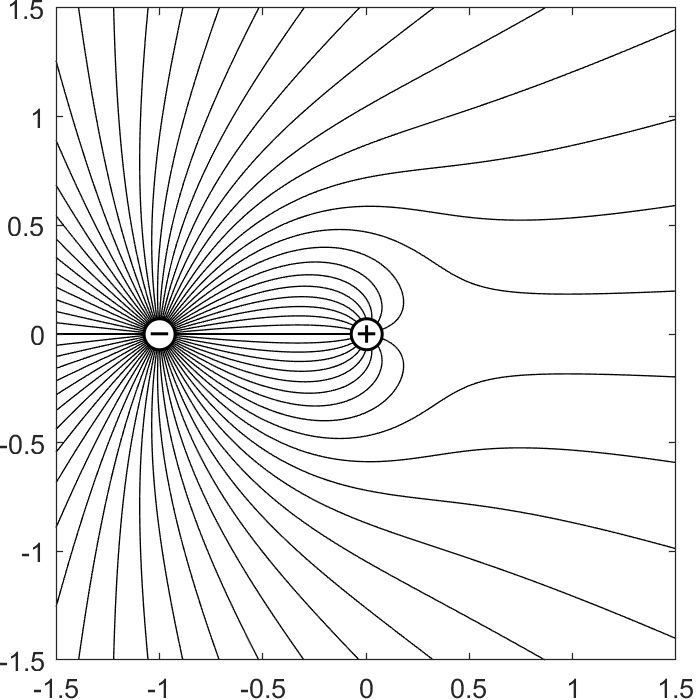}
		\caption{}
		\label{sub_dipole_streamlines}
	\end{subfigure}
	\begin{subfigure}{0.45\textwidth}
		\includegraphics[width = \textwidth]{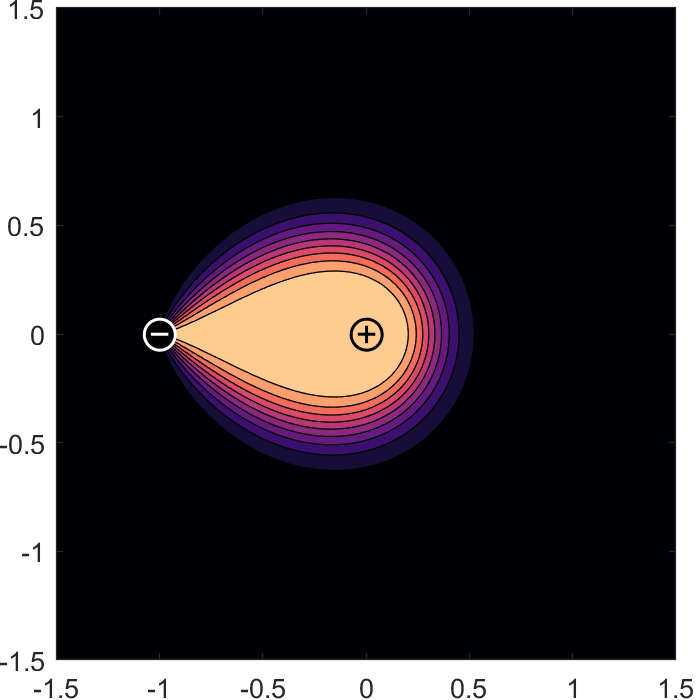}
		\caption{}
		\label{sub_dipole_concentration}
	\end{subfigure}
	\caption{Streamlines (\subref{sub_dipole_streamlines}) and concentration profile (\subref{sub_dipole_concentration}) for a finite dipole with $\alpha = 4$ and $\Pen = 10$}
	\label{figure_dipole}

\end{figure}

The simplest case of a multipolar flow for which a steady-state concentration profile exists is the finite dipole, illustrated in figure \ref{figure_dipole}, consisting of a single injection of unit flow rate and a single aspiration of flow rate $\alpha > 1$, separated by a finite distance. Flow and concentration in the finite dipole are exactly equivalent to those in the microfluidic probe \citep{juncker2005multipurpose}, and provide the starting point for our analysis. The flow field and an example of a concentration field in the finite dipole are illustrated in figure \ref{figure_dipole}. The complex potential in a finite dipole is given by

\begin{equation} \label{dipole_flow}
	\Phi = \text{log}\left( z \right) - \alpha \ \text{log} \left( z + 1 \right)
\end{equation}

In cylindrical coordinates, the convection-diffusion equation for the dipole flow becomes

\begin{dmath}
	\frac{\partial^2 c}{\partial r^2} + \frac{1}{r}\frac{\partial c}{\partial r} + \frac{1}{r^2} \frac{\partial^2 c}{\partial \theta^2} = - \Pen \left( \frac{1}{r \left( r^2 + 2 r \text{cos} \theta + 1 \right) } \left( \left( \left( \alpha - 1 \right) r^2 + \left( \alpha - 2 \right) r \text{cos}\theta - 1 \right) \frac{\partial c}{\partial r} - \alpha \ \text{sin} \theta \frac{\partial c}{\partial \theta} \right) \right)
\end{dmath}

The concentration is fixed a $c = 1$ for the injection aperture, $c = 0$ at infinity and zero flux at the aspiration aperture. Several approaches exist to solve this equation. The classic method is to use perturbation methods to analyze the behavior of $c$ in the limits of very high or very low $\Pen$. This problem has to be solved using singular perturbations, similar to the approach used by Acrivos and Taylor to analyze convection-diffusion around a spherical obstacle \citep{acrivos1962heat}. This is feasible when $\alpha = 2$ and the boundary layer separating the region of high concentration from the region of null concentration takes the form of a circle of radius 1. However, for general values of $\alpha$, the boundary layer’s position is not trivial, which makes it hard to obtain a solution in terms of singular perturbations. Another solution avenue is to use material surface tracking, which is often used in groundwater problems involving point sources and sinks \citep{bodvarsson1972thermal}. However, this method usually neglects the effects of lateral diffusion, which we want to account for in our analysis.

\subsection{The problem in streamline coordinates}

To simplify the problem, we use a transformation to the Boussinesq coordinates \citep{boussinesq1902pouvoir}. Since the system of equations \ref{laplace_eq} and \ref{convection_diffusion_eq} is conformally invariant, we can use the complex potential $\Phi$ as a map from the domain $z = x+i y$ of dipolar flow to the streamline coordinate plane $\Phi = \phi + i \psi$ where the flow is a simple, constant plane flow. This is similar to the hodograph method used in interface problems for groundwater flows \citep{bear1964some} \citep{de1965many} \citep{strack1972some}. In the streamline coordinate domain, the convection-diffusion equation is reduced to the simpler form

\begin{equation}\label{convection_diffusion_streamline}
	\frac{\partial^2 c}{\partial \psi^2} + \frac{\partial^2 c}{\partial \phi^2} - \Pen \frac{\partial c}{\partial \phi} = 0
\end{equation}

This represents convection-diffusion in a uniform plane flow. There remains to impose the proper boundary conditions on \ref{convection_diffusion_streamline}. To determine these boundary conditions, we must determine where the important features of the dipole flow map in the streamline coordinate domain. These important features are the symmetry axis at $x = 0$, split in 4 different segments by the two apertures and the stagnation point, as well as the streamline going from the stagnation point to the aspiration aperture. The map of each segment in streamline coordinate can easily be obtained by using the complex potential to map each endpoint. By the definition of our map, any segment of a streamline maps to a straight segment of constant $\psi$ in the $\Phi$ domain. These features are illustrated in figure \ref{figure_streamline} and compiled in table \ref{table_streamline}.

The equivalent problem to solve is thus the problem of convection-diffusion inside of a strip with zero-flux (symmetry) conditions on the top and bottom walls. In addition, the strip is separated by a semi-infinite segment of zero flux around $\psi = 0$, which separates a region for which $c \rightarrow 0$ as $\phi \rightarrow -\infty$ and one for which $c \rightarrow 1$ as $\phi \rightarrow -\infty$. This semi-infinite boundary condition makes the problem particularly hard to solve analytically, even if the underlying equation is relatively simple. We will give solutions to this problem in the limits of high and low Peclet numbers, and use these solutions to obtain concentration profiles for different multipole geometries.

\begin{figure}
	\centering
	\begin{subfigure}{0.45\textwidth}
		\includegraphics[width=\textwidth]{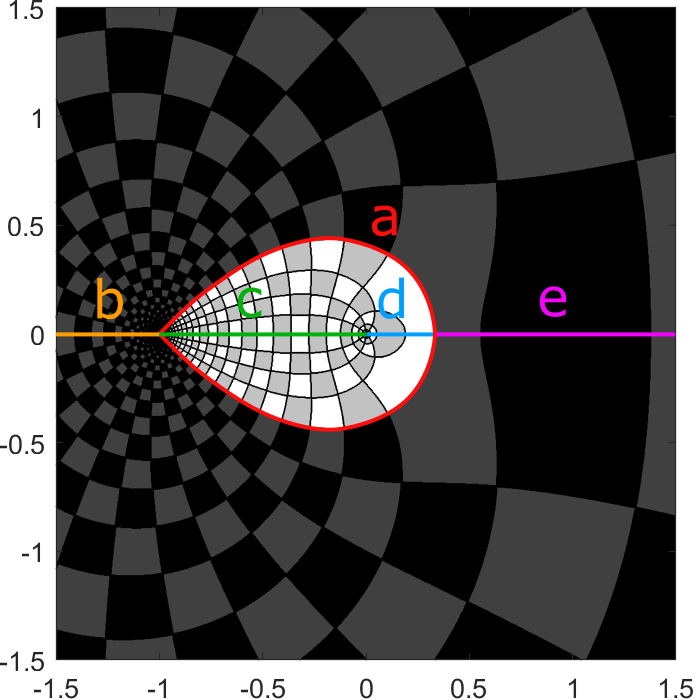}
		\caption{}
		\label{figure_grid_dipole}
	\end{subfigure}
	\begin{subfigure}{0.45\textwidth}
		\includegraphics[width=\textwidth]{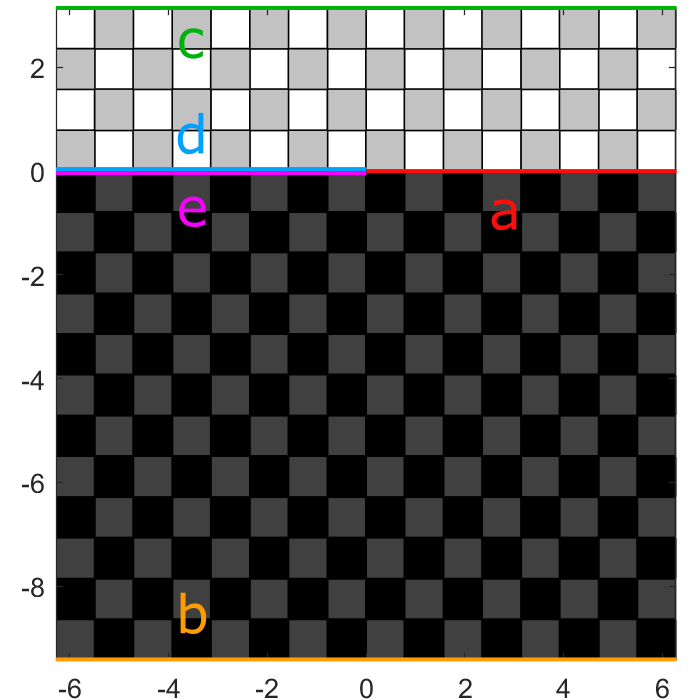}
		\caption{}
		\label{figure_flat_grid_dipole}
	\end{subfigure}

	\caption{Map of the important features of the finite dipole domain (\subref{figure_grid_dipole}) and their equivalent in the streamline coordinate domain (\subref{figure_flat_grid_dipole})}
	\label{figure_streamline}
\end{figure}

\begin{table}
  \begin{center}
  \begin{tabular}{cccc}
	\hline
	Segment 	&	Z-plane coordinates 	& Streamline coordinates 	& 	Boundary condition \\ \hline
 	a)			& $r = \frac{\text{sin}\left( \nicefrac{1}{\alpha} \ \theta \right)}{\text{sin} \left( \left( 1 - \nicefrac{1}{\alpha} \right) \theta \right)}$					& $\phi \ \in \ ] 0, \infty[$		& Continuity 	\\
				& see \citep{goyette2019microfluidic}	& $\psi = 0$	& 	\\ \hline
	b)			& $x \ \in \ ] -\infty, 1 [$		& $\phi \ \in \ ] -\infty,\infty[$	& Symmetry 	\\
				& $y = 0$						& $\psi = \left(1 - \alpha \right) \pi$	&  	\\ \hline
	c)			& $x \ \in \ ]-1, 0[$				& $\phi \ \in \ ]-\infty, \infty[$	& Symmetry 	\\
				& $y = 0$						& $\psi = \pi$				&  	\\ \hline
	d)			& $x \ \in \ ]0, x_{stag}[$		& $\phi \ \in \ ]-\infty,0[$		& Symmetry 	\\
				& $y = 0$						& $\psi = 0^+$				&  	\\ \hline
	e)			& $x \ \in \ ]x_{stag},\infty[$	& $\phi \ \in \ ]-\infty,0[$		& Symmetry 	\\
				& $y = 0$						& $\psi = 0^-$				&  	\\ \hline
  \end{tabular}
  \caption{Important features of the dipole domain and their equivalent in the streamline coordinate domain. See figure \ref{figure_streamline} for correspondence. Z-plane expression for segment a) is expressed in polar coordinates centered on the injection aperture for simplicity}
  \label{table_streamline}
  \end{center}
\end{table}

\subsection{Position of the stagnation point}

In the streamline coordinate domain, there is a semi-infinite strip of zero flux at $\psi = 0$ going from $\psi \rightarrow$ to $\phi = \phi_{stag}$, where $\phi_{stag}$ is the image of the stagnation point in the dipole flow. The stagnation point represents the point in the dipole flow domain where the velocity field vanishes. It can be shown that is position is given by

\begin{equation}
	z_{stag} = \frac{1}{\alpha - 1}
\end{equation}

Its image in the streamline domain is thus given by

\begin{equation}
	\Phi_{stag} = \text{log} \left( \frac{1}{\alpha - 1} \right) - \alpha \ \text{log} \left( \frac{\alpha}{\alpha - 1} \right)
\end{equation}

For the remainder of the text, we will use $\psi + i \phi = \Phi + \Phi_{stag}$ as a transformation, so that the semi-infinite boundary ends at the origin in the streamline domain.

\subsection{Infinite Peclet limit}


The first and easiest approximation that can be made is the approximation of infinite Peclet number, which corresponds to the complete neglect of lateral diffusion. In the limit of infinite Peclet number, the convection-diffusion equation \ref{convection_diffusion_eq} becomes

\begin{equation}
	\nabla \phi \cdot \nabla c = 0
\end{equation}

Which has the general solution

\begin{equation} \label{infinite_pe_solution}
	c = f \left( \psi \right)
\end{equation}

Where $f$ is a function to be determined from the boundary conditions. In many transport problems, \ref{infinite_pe_solution} is unable to fulfill all boundary conditions, and a complete solution has to be determined using boundary layer methods \citep{leal2007advanced}. However, in the case of the finite dipole, the pure convective solution is a valid first approximation for the concentration profile. Plugging in the boundary conditions, we obtain the solution

\begin{equation} \label{infinite_pe_solution_dipole}
	c \left( \psi \right) =
	\begin{cases} 
		1 & \psi \geq 0 \\
		0 & \psi < 0
   \end{cases}
\end{equation}

It is also possible, as we have shown elsewhere \citep{goyette2019microfluidic}, to express the separating line in polar coordinates as a ratio of sine functions. This allows us to rewrite solution \ref{infinite_pe_solution_dipole} as

\begin{equation}
	c \left( r, \theta \right) =
	\begin{cases}
		1 & r \leq \frac{\text{sin} \left( \nicefrac{1}{\alpha} \ \theta \right)}{\text{sin} \left( \left( 1 - \nicefrac{1}{\alpha} \right) \theta \right)} \\
		0 & \text{elsewhere}
	\end{cases}
\end{equation}

Where the origin is taken at the injection aperture.

\subsection{Approximation for large Peclet}

We now turn to the high (but not infinite) $\Pen$ limit. The semi-infinite boundary condition at $\psi = 0$ seems to suggest that we could use the Wiener-Hopf method to solve this problem, as in \citep{springer1973solution} \citep{springer1974solution}\citep{carlson1947reflection}. However, the addition of symmetry conditions at $\psi = \pi$ and $\psi = \left(1 - \alpha \right) \pi$, as well
as the different limit conditions for $c$ as $\phi \rightarrow -\infty$ makes the problem difficult to pose using this formalism. We thus look for a simplified formulation of the problem that still captures the important features of the solution. To do so, we first observe the behavior of the solution near $\Phi = 0$, that is near the edge of the semi-infinite no-flux boundary. At high $\Pen$, there is a region around this point where the effect of the boundaries at $\psi = \pi$ and $\psi = \left( 1 - \alpha \right) \pi$ are negligible. A local solution valid in this region can be obtained by solving the simplified boundary value problem consisting of equation \ref{convection_diffusion_streamline} with the boundary conditions

\begin{equation}
	\begin{aligned}
		\frac{\partial c}{\partial \psi} = 0 \quad & \phi < 0, \psi = 0 \\
		c = 1 \quad & \phi \rightarrow -\infty, \psi > 0 \\
		c = 0 \quad & \phi \rightarrow -\infty, \psi \leq 0 \\ 
	\end{aligned}
\end{equation}

By symmetry of the problem, we can deduce that $c$ will be equal to $\nicefrac{1}{2}$ for $\psi \geq 0, \phi = 0$. This allows us to independently solve both halves of the plane. Specifically, the problem in both halves becomes identical to the leading-edge problem from \cite{cummings1999two}. We can thus apply the solution directly and obtain the complete concentration profile as a piecewise solution. The solution obtained is continuous on the entire domain apart from the semi-infinite BC.

\begin{equation} \label{solution_wake}
	c \left( \psi, \phi \right) =
	\begin{cases} 
      \nicefrac{1}{2} \left( 1 - \text{erf} \left( \text{Im} \sqrt{ \Pen \left( \Phi \right) } \right) \right) & \psi < 0 \\
      \nicefrac{1}{2} \left( 1 + \text{erf} \left( \text{Im} \sqrt{ \Pen \left( \Phi \right) } \right) \right) & \psi \geq 0 \\
   \end{cases}
\end{equation}

\begin{figure}
	\centering
	\begin{subfigure}{0.45\textwidth}
		\includegraphics[width=\textwidth]{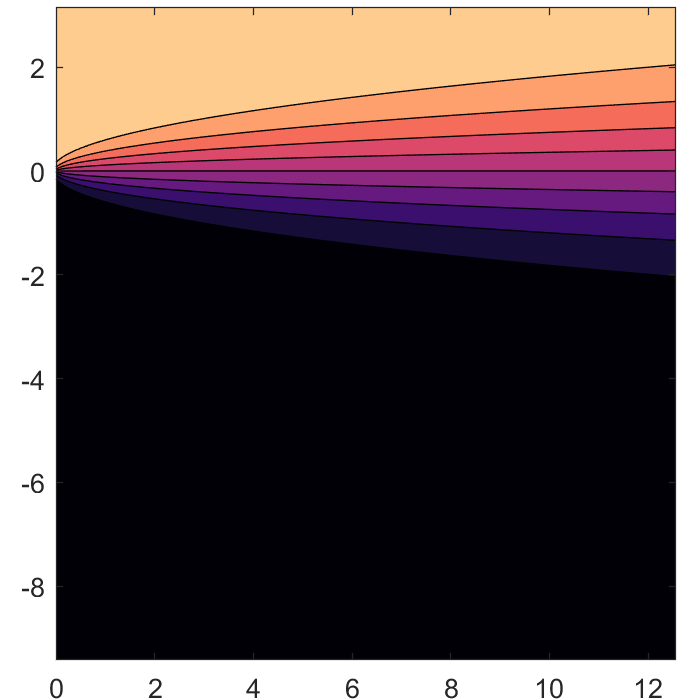}
		\caption{}
		\label{figure_wake}
	\end{subfigure}
	\begin{subfigure}{0.45\textwidth}
		\includegraphics[width=\textwidth]{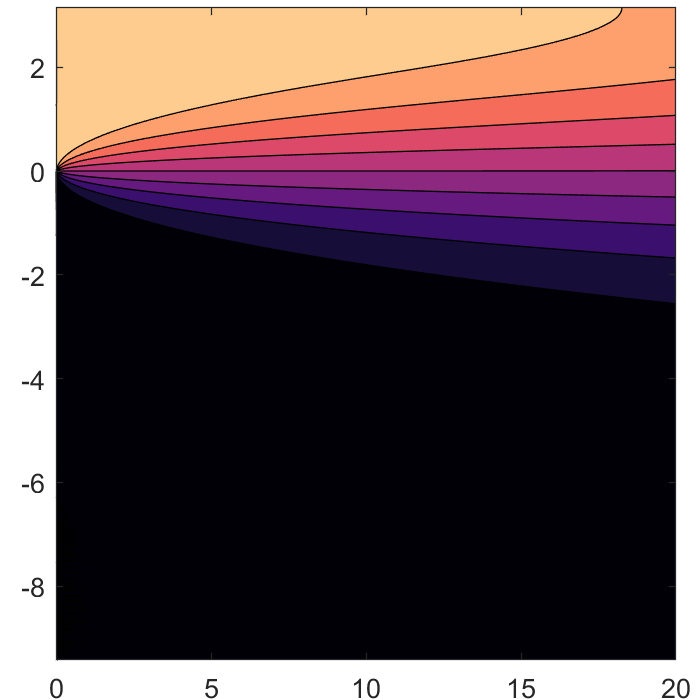}
		\caption{}
		\label{figure_helmholtz}
	\end{subfigure}

	\caption{Streamline coordinate solutions for the finite dipole at high $\Pen$. \subref{figure_wake}) Solution \ref{solution_wake}, valid around the stagnation point and breaking down far downstream and \subref{figure_helmholtz}) Solution \ref{solution_helmholtz} valid far downstream but breaking down near the stagnation point}
	\label{figure_highPe}
\end{figure}

This solution is illustrated in figure \ref{figure_wake}

\subsubsection{Breakdown of the approximation}

The solution \ref{solution_wake} is valid near the origin and for high values of $\Pen$, however it breaks down for sufficiently large values of $\phi$. This can be seen intuitively by observing that very far downstream, the solution yields a uniform concentration of $c = \nicefrac{1}{2}$ while a simple firstprinciple analysis of the original problem shows that far enough down the channel, the concentration must be $c = \nicefrac{1}{\alpha}$ everywhere. More specifically, we have neglected the effect of the symmetry conditions at $\psi = \pi$ and $\psi = \left( 1 - \alpha \right) \pi$, which don’t influence the wake near the mixing region, but eventually must be considered sufficiently far downstream. We have previously shown \cite{goyette2019microfluidic} that this breakdown region happens when $\psi > O \left( \Pen \right)$. When the solution is transformed back to the dipole domain, the breakdown region where the solution is no longer valid is confined to a circle or radius $O \left( e^{-\Pen} \right)$ around the aspiration aperture.

\subsection{Second approximation for large Peclet} \label{section_high_Pe_2}

In some cases, we may be interested in solutions for the highly convective regime in which we need accuracy far downstream. In this case, we can use different approximations for the problem in streamline coordinates and obtain a second solution. For very high values of $\Pen$, we can be justified in assuming that no mixing of the fluid happens before $\phi > \phi_{stag}$. In that case, we can assume that concentration will be constant in each channel region for $\phi < \phi_{stag}$. We can then solve equation \ref{convection_diffusion_streamline} for the semi-infinite strip $\phi > 0$ , $\psi \in ] \left( 1 - \alpha \right) \pi, \pi [$.

Using the change of variable $c \left( \phi, \psi \right) = \text{exp} \left( \nicefrac{1}{2} \Pen \ \phi \right) U \left( \phi, \psi \right)$, we can bring \ref{convection_diffusion_streamline} to the form of the helmholtz equation

\begin{equation}
	\frac{\partial^2 U}{\partial \psi^2} + \frac{\partial^2 U}{\partial \phi^2} = \frac{1}{4} \Pen^2 U
\end{equation}

The Green's function for this boundary value problem is known and is given by \cite{polyanin2015handbook}

\begin{dmath}
	G \left( \phi, \psi, \phi', \psi' \right) = \frac{1}{2 \alpha \pi} \sum_{n=0}^\infty \frac{\epsilon_n}{\beta_n} \left( \text{exp}\left( -\beta_n \abs{\phi - \phi'} \right) - \text{exp}\left( -\beta_n \abs{\phi + \phi'} \right) \right) \text{cos} \left( \frac{n}{\alpha} \left( \psi - \pi \right) \right) \text{cos} \left( \frac{n}{\alpha} \psi' \right)
\end{dmath}

With $\beta_n = \sqrt{ \frac{n^2}{\alpha^2} + \frac{\Pen^2}{4}}$, $\epsilon_0 = 2$ and $\epsilon_n = 1$ for $n \neq 0$.  This Green’s function can be convoluted with the value of $c$ at $\psi = 0$ to obtain the concentration profile. Since we approximate the concentration profile at $\psi = 0$ to be simply a step function, we obtain

\begin{equation}
	c \left( \phi, \psi \right) = \int_0^{\pi}  \frac{\partial G}{\partial \phi'}\left( \phi, \psi, \phi', \psi' \right) \rvert_{\phi' = 0} \ d \psi'
\end{equation}

Which can be simplified to

\begin{equation} \label{solution_helmholtz}
	c \left( \phi, \psi \right) = \frac{1}{\alpha} + \frac{2}{\pi} \sum_{n=1}^{\infty} \frac{1}{n} \text{exp} \left( \left( \frac{\Pen}{2} - \sqrt{\frac{n^2}{\alpha^2} + \frac{\Pen^2}{4} } \right) \phi \right) \text{sin}\left( \frac{\pi n}{\alpha} \right) \text{cos} \left( \frac{n}{\alpha} \left( \psi - \pi \right) \right)
\end{equation}

About 10 terms are required to have an error of less than $10^{-3}$ everywhere except near the branch cut at $\phi = 0$, where the solution is already not valid. Adding further terms beyond that only resolves the region around $\Phi = 0$ more. This solution, illustrated in figure \ref{figure_helmholtz} breaks down in a region of order $O \left( \sqrt{\Pen} \right)$ around the branch cut since it neglects retrodiffusion in that region.

\subsection{Error for each approximation}

\begin{figure}
	\centering
	\begin{subfigure}{0.45\textwidth}
		\includegraphics[width=\textwidth]{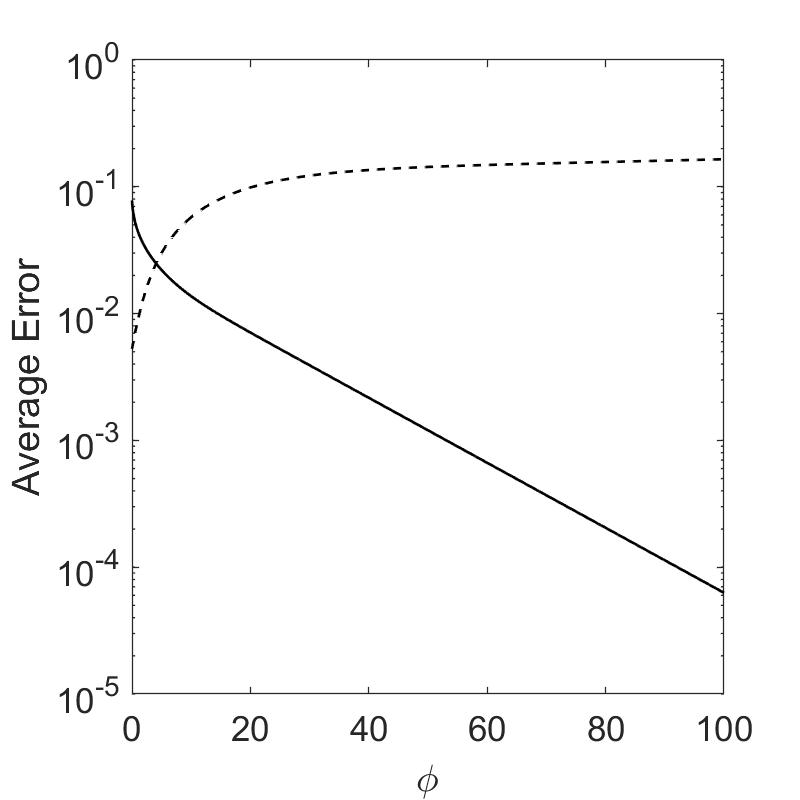}
		\caption{$\Pen = 1$}
		\label{fig_err_Pe1}
	\end{subfigure}
	\begin{subfigure}{0.45\textwidth}
		\includegraphics[width=\textwidth]{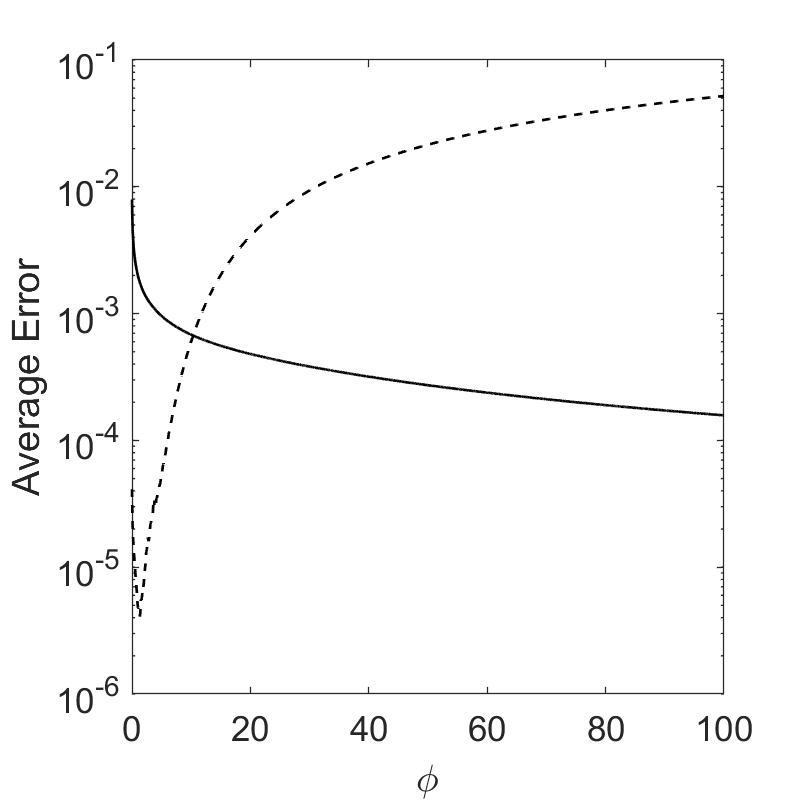}
		\caption{$\Pen = 10$}
		\label{fig_err_Pe10}
	\end{subfigure}
	\begin{subfigure}{0.45\textwidth}
		\includegraphics[width=\textwidth]{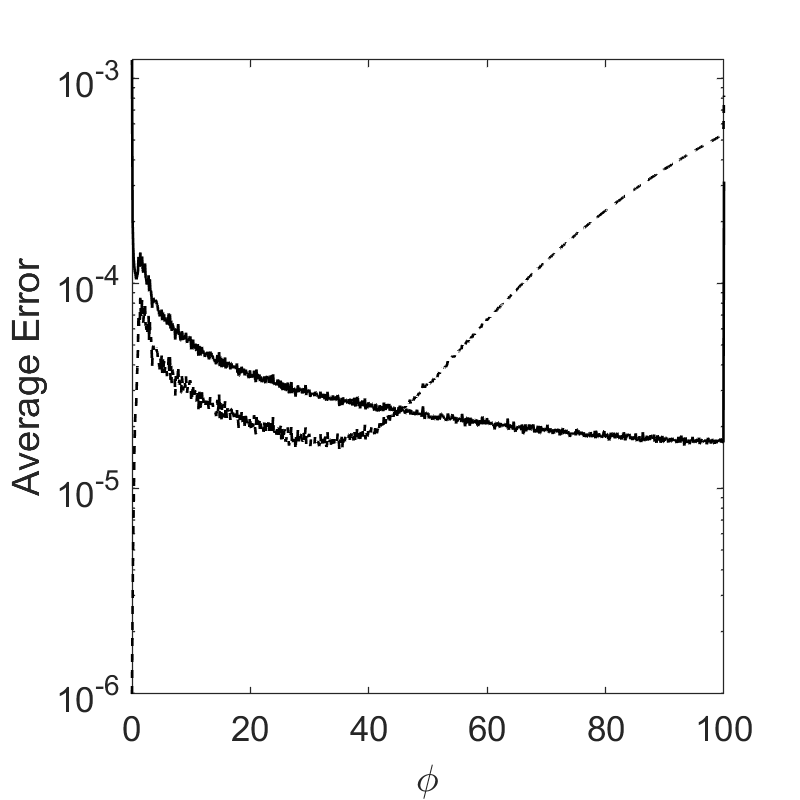}
		\caption{$\Pen = 100$}
		\label{fig_err_Pe100}
	\end{subfigure}
	\begin{subfigure}{0.45\textwidth}
		\includegraphics[width=\textwidth]{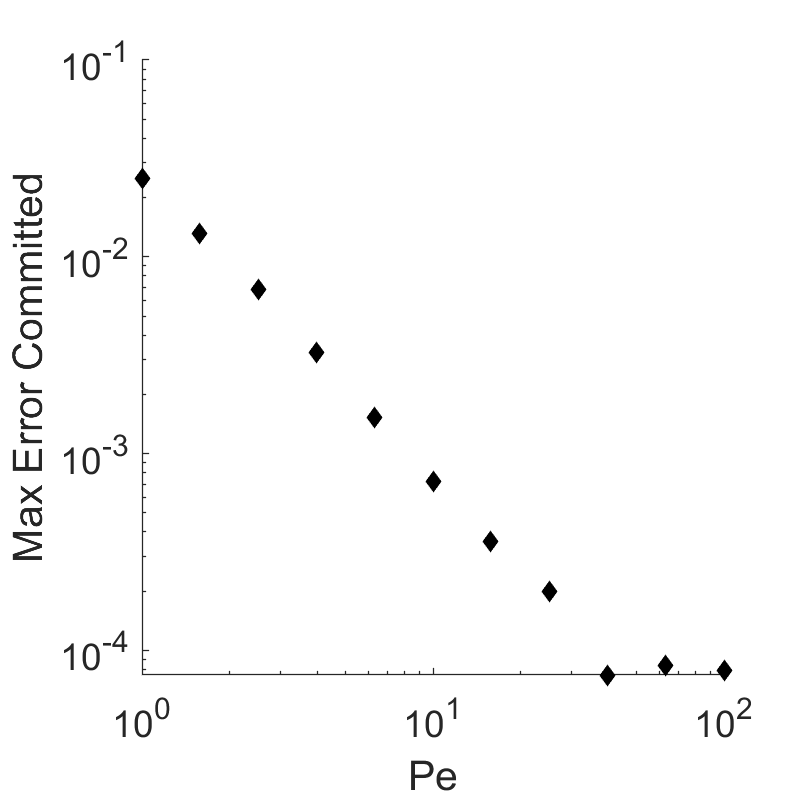}
		\caption{Maximum error committed}
		\label{fig_max_err}
	\end{subfigure}

	\caption{Error averaged along $\psi$ axis for different values of $\Pen$ for the self-similar solution (dashed) and series solution (full line) for $\alpha = 4$ and different values of $\Pen$. Subfigure \subref{fig_max_err} shows the maximum error committed by combining both approximations. The error for each graph was obtained by comparing with 2D finite element simulations of the streamline problem done in COMSOL multiphysics. For each value of $\phi$, the error was integrated along the $\psi$ axis}
	\label{figure_compare}
\end{figure}

We have obtained two different approximations of the concentration profile in streamline coordinates for the dipole problem. Both break down in different regions and in general, we will use the one which is valid in the region of interest, depending on the problem. For multipolar devices with hydrodynamic flow confinement, solution \ref{solution_wake} is perfectly appropriate, as the breakdown region will be confined to an exponentially small region under the aspiration aperture. If however, we want the concentration profile for non
hydrodynamically confined devices such as impinging flows (presented in section \ref{section_impinging}), we may use the second solution, valid far enough downstream. In figure \ref{figure_compare}, we compare both solutions to finite element simulations to determine the error along the $\phi$ axis. We notice that the value of $\phi$ for which the error is larger for the self-similar solution is of the order $O \left( \Pen \right)$. We also note that by using a combination of the two solutions, we can use the high $\Pen$ solution for values of $\Pen$ as low as 1 while keeping an average error that is smaller than 2 \%. The maximum error committed decreases approximately with $\Pen^{-1.5}$, as shown in figure \ref{fig_max_err}.

\subsection{Approximation for very low Peclet}

In the limit of very low Peclet number, we may approximate that diffusion in the $\psi$ direction happens much faster than convection, and that at any point in the streamline domain the fluid is completely mixed along any vertical line so that $c$ is only a function of $phi$. This approximation allows us to separate the streamline domain in 3 1-dimensional subdomains and solve a simplified equation for each.

The first subdomain we analyze is the part of the strip defined by $\phi \geq 0$, corresponding to the region after the separating cut of zero derivative. In the very low Peclet regime, we can approximate that this region is perfectly mixed and has constant concentration
everywhere. The concentration in this region will be the equilibrium concentration, that is $c = \nicefrac{1}{\alpha}$. As $\Pen$ increases, this approximation will break down in a region around $\Phi = 0$, where mixing will not yet be complete.

Once we have fixed the concentration for $\phi \geq 0$ to a constant $c = \nicefrac{1}{\alpha}$, the top and bottom half of the strip for $\phi < 0$ can be independently solved as separate 1-dimensional problems. In both cases, we are solving the 1-D convection-diffusion equation

\begin{equation}
	\frac{\partial^2c}{\partial \phi^2} - \Pen \frac{\partial c}{\partial \phi} = 0
\end{equation}

With $c = \nicefrac{1}{\alpha}$ at $\phi = 0$ and concentration at $\phi = -\infty$ fixed at either 0 (top half-strip) or 1 (bottom half-strip). This gives us a piecewise definition for the streamline coordinates solution at very low Peclet numbers

\begin{equation}
	c \left( \phi, \psi \right) =
	\begin{cases} 
      \frac{1}{\alpha} & \phi \geq 0 \\
      \frac{1}{\alpha} e^{\Pen \phi} & \phi < 0 \ \& \ \psi > 0 \\
      1 - \frac{\alpha - 1}{\alpha} e^{\Pen \phi} & \phi < 0 \ \& \ \psi < 0 \\
   \end{cases}
\end{equation}

This solution can be taken back to the dipole flow domain using the complex potential as a transformation, as was done for the high $\Pen$ solution.

\section{Transforming the dipole solution}

\begin{figure}

	\centering
	\begin{subfigure}{0.3\textwidth}
		\includegraphics[width=\textwidth]{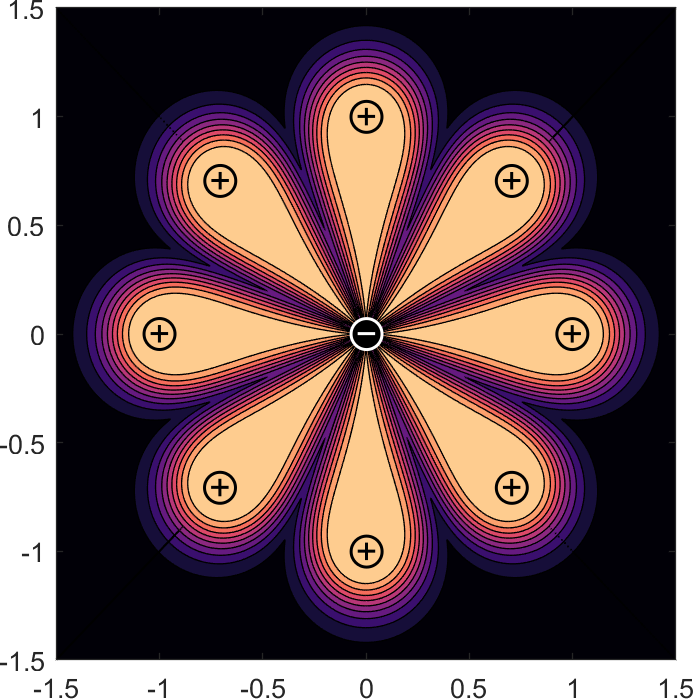}
		\caption{$z = z^{'8} - 1$}
	\end{subfigure}
	\begin{subfigure}{0.3\textwidth}
		\includegraphics[width=\textwidth]{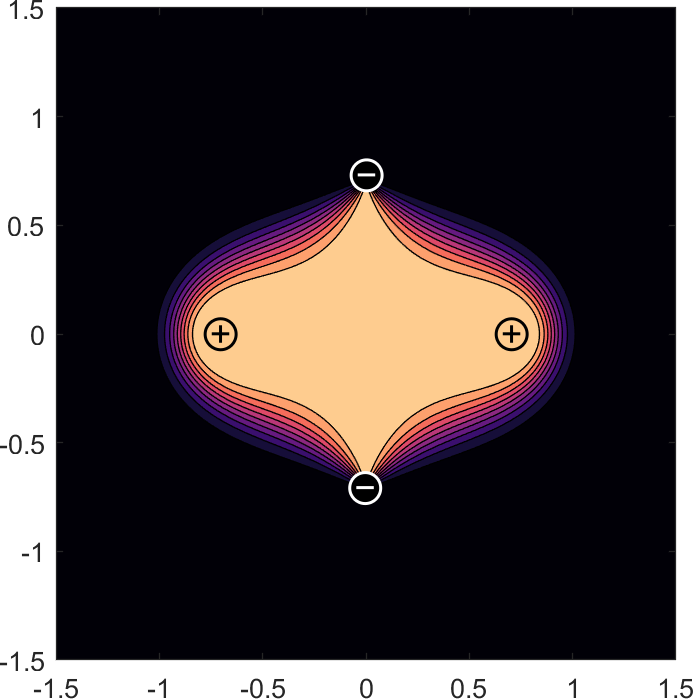}
		\caption{$z = z^{'2} - 0.5$}
	\end{subfigure}
	\begin{subfigure}{0.3\textwidth}
		\includegraphics[width=\textwidth]{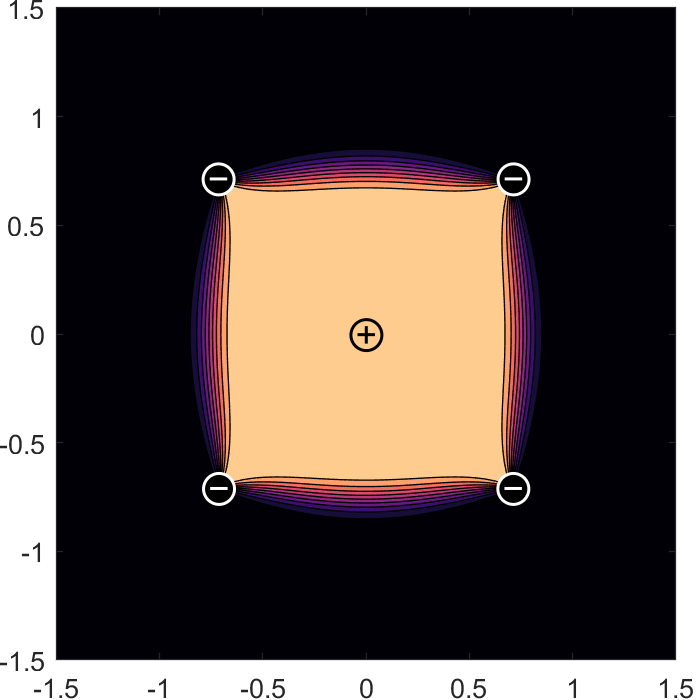}
		\caption{$z = z^{'4}$}
	\end{subfigure}
	\caption{Concentration profiles obtained using combinations of translations and power tranforms}
	\label{figure_transformed}
\end{figure}

The dipole solution, be it for high or low values of $\Pen$, can be used as a starting point to generate more elaborate solutions without needing to solve new transport problems. We can once again exploit the fact that the plane convection-diffusion equation and the Laplace equation describing the flow field form a conformally invariant system of PDEs. This allows us to use simple conformal maps to map the dipole flow to other multipolar flows with more apertures and obtain solution of the transport problem in those flows directly. Examples of simple transforms that can be used include power transforms for generating rotationally symmetrical geometries or inversion transforms to obtain nonconfined flow profiles, as will be described in the following sections. A selection of concentration profiles obtained from the dipole solution are illustrated in figure \ref{figure_transformed}.

\section{The impinging flows solution}\label{section_impinging}

The finite dipole solution can be transformed to obtain concentration profiles in geometries without hydrodynamic confinement. For instance, starting from the solution \ref{solution_wake} for the dipole flow geometry \ref{dipole_flow}, we can use the transformation

\begin{equation}
	z = z^{'-1} - 1
\end{equation}

To obtain the concentration in a flow caracterized by an injection aperture of rate $\beta = \alpha - 1$ rate and null concentration at the origin and another injection aperture of unit rate and unit concentration at $z = 1$. The flow profile for the impinging flows is given by the gradient of the potential function

\begin{equation}
	\Phi \left( z \right) = \text{log} \left( z - 1 \right) + \beta \ \text{log} \left( z \right)
\end{equation}

\begin{figure}
	\centering
	\begin{subfigure}{0.3\textwidth}
		\includegraphics[width=\textwidth]{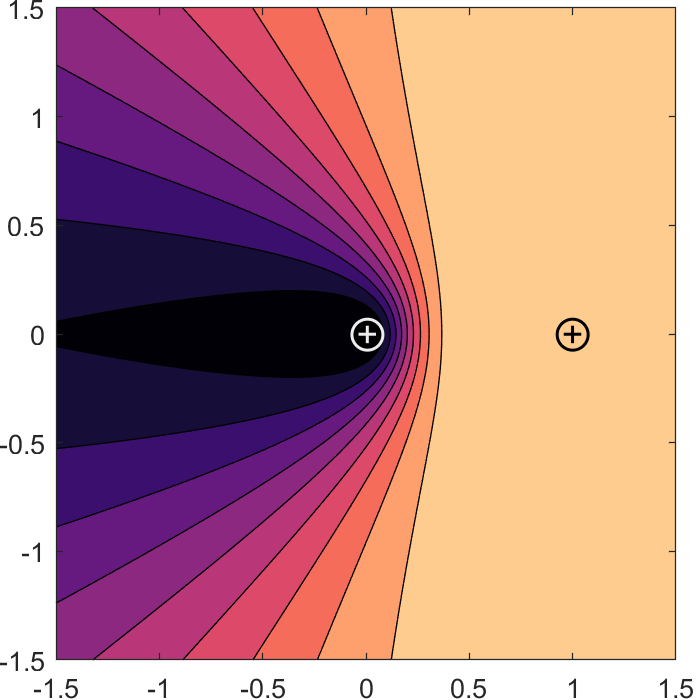}
		\caption{$\beta = \nicefrac{1}{4}$}
	\end{subfigure}
	\begin{subfigure}{0.3\textwidth}
		\includegraphics[width=\textwidth]{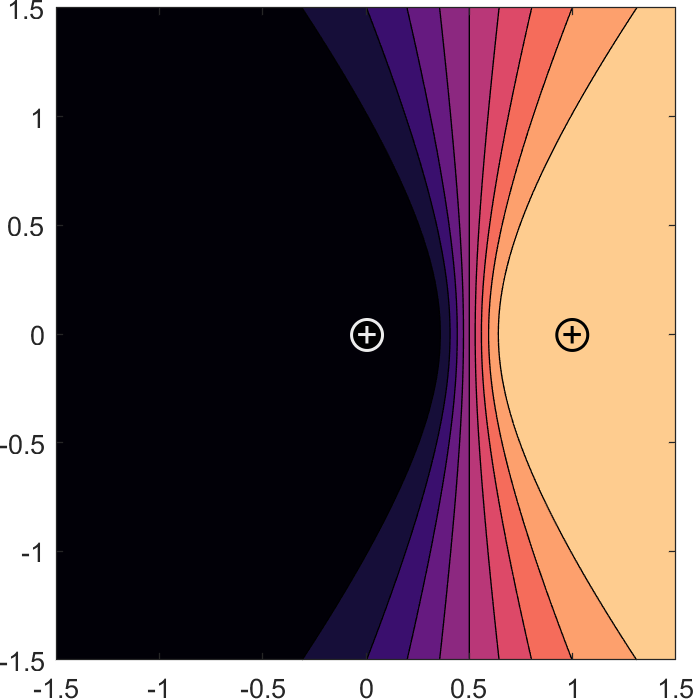}
		\caption{$\beta = 1$}
	\end{subfigure}
	\begin{subfigure}{0.3\textwidth}
		\includegraphics[width=\textwidth]{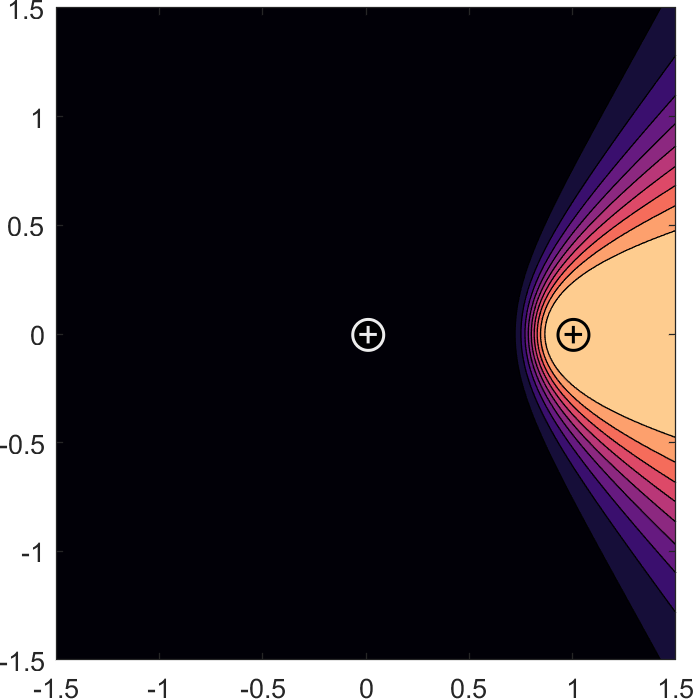}
		\caption{$\beta = 4$}
	\end{subfigure}
	
	\caption{Solutions for impinging flows at $\Pen = 10$ for various values of $\beta$}
	\label{figure_impinging}
\end{figure}

In the case of $\beta = 1$, the two singularities inject at the same rate, and a vertical line of concentration $c = \nicefrac{1}{2}$ appears around $z = \nicefrac{1}{2}$. Very close to this stagnation point, the flow behaves like a purely extensional flow and the concentration profile approaches the well-known y-independent solution \citep{bazant2004conformal} \citep{qasaimeh2011microfluidic}. However, as $\abs{y}$ increases, the velocity decreases and a broadening of the concentration gradient is observed. In the case of $\beta > 1$, a hyperbolic wake is formed around the aperture at $z = 1$, and a similar wake is observed around $z = 0$ when $\beta < 1$. Different solutions for the impinging flows are illustrated at figure \ref{figure_impinging}.

This solution can be used to model transport from two jets of different temperature impinging on a 2D plate, or temperature in an aquifer with two different cylindrical injection wells. In microfluidics, impinging flow devices can be used to generate long adjustable concentration gradients in an open-space context.

\section{Other starting geometries}

\begin{figure}
	\centering
	\begin{subfigure}{0.3\textwidth}
		\includegraphics[width=\textwidth]{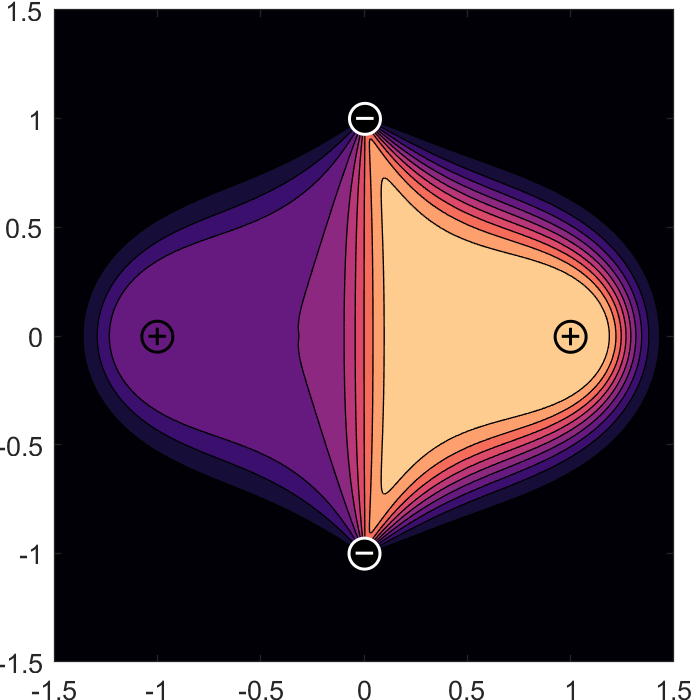}
	\end{subfigure}
	\begin{subfigure}{0.3\textwidth}
		\includegraphics[width=\textwidth]{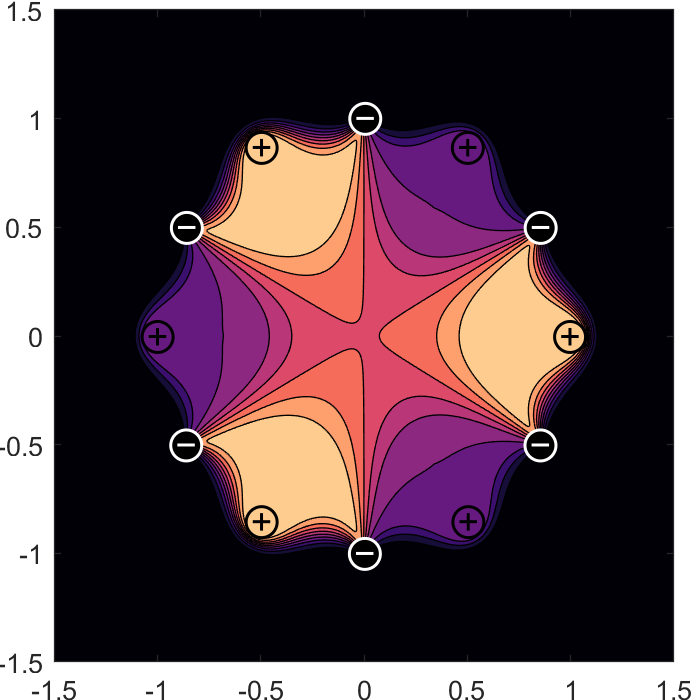}
	\end{subfigure}
	\begin{subfigure}{0.3\textwidth}
		\includegraphics[width=\textwidth]{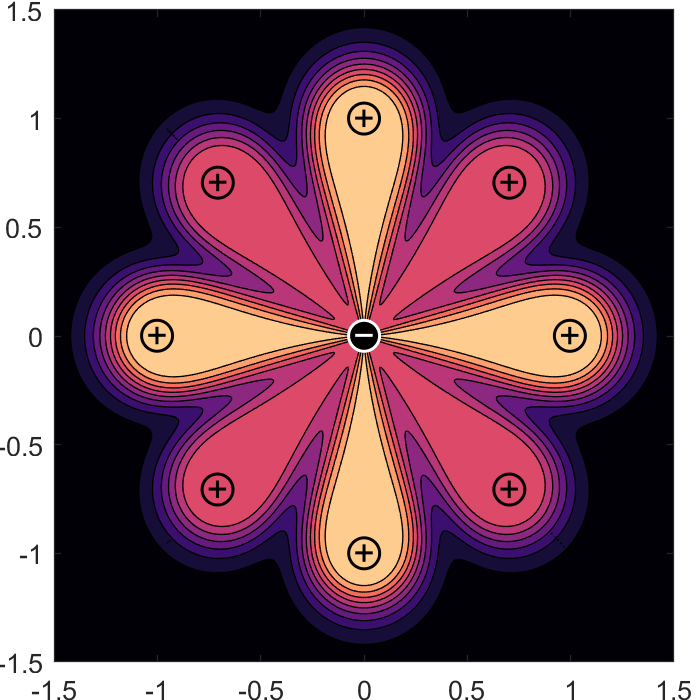}
	\end{subfigure}
	
	\caption{Concentration profiles for multipolar geometries not obtainable from the simple dipole solution}
	\label{figure_other_geometries}
\end{figure}

In the previous sections, we have been exploiting the solution to the dipole geometry problem described in \ref{section_dipole}, and finding derived solutions by using conformal transforms. However, the method outlined here is quite general and can easily be extended to more
complex starting geometries. In cases when the arrangement of sources and sinks cannot be obtained by transforming an already known solution, we have to start by mapping the problem to streamline coordinates and see how the branch cuts map in the channel
geometry. Once the problem is properly posed in the streamline coordinate domain, it can be solved using appropriate approximations and then mapped back to the multipole flow. This includes, but is not limited to, applications where there are injection apertures injecting more than a single reagent or reagent concentration. Examples of solutions obtained from different starting geometries are illustrated in figure \ref{figure_other_geometries}.

\subsection{Thermal plume in straight flow}

One example of a different starting geometry is the case of a thermal plume in a straight flow. This is generated by a single injection aperture at the origin in a uniform plane flow. This geometry can be found in problems of injection of cold water in aquifers, for example waste from groundwater heat pumps or in the analysis of the contamination of underground water sources. This arrangement cannot trivially be obtained by applying simple transforms to the dipole solution presented in the previous sections. However, we can apply the same reasoning as we did in section \ref{section_dipole} to find the form of the problem in streamline coordinates. The obtained streamline coordinate problem is very similar to the finite dipole one, and the same solution procedure can be applied.

Flow in the thermal plume arrangement is described by the complex potential

\begin{equation}
	\Phi = z + \alpha \ \text{log} \left( z \right)
\end{equation}

Which corresponds to a dipole source of unit strength at infinity and an injection aperture of rate $\alpha$ at the origin. This flow possesses a single stagnation point at $z_{stag} = - \alpha$. This stagnation point transforms to the point $\Phi_{stag}$ given by

\begin{equation}
	\Phi_{stag} = - \alpha + \alpha \ \text{log} \left( - \alpha \right) = \alpha \left( \text{log} \left( \alpha \right) - 1 \right) + i \ \alpha \ \pi
\end{equation}

\begin{figure}
	\centering
	\begin{subfigure}{0.45\textwidth}
		\includegraphics[width=\textwidth]{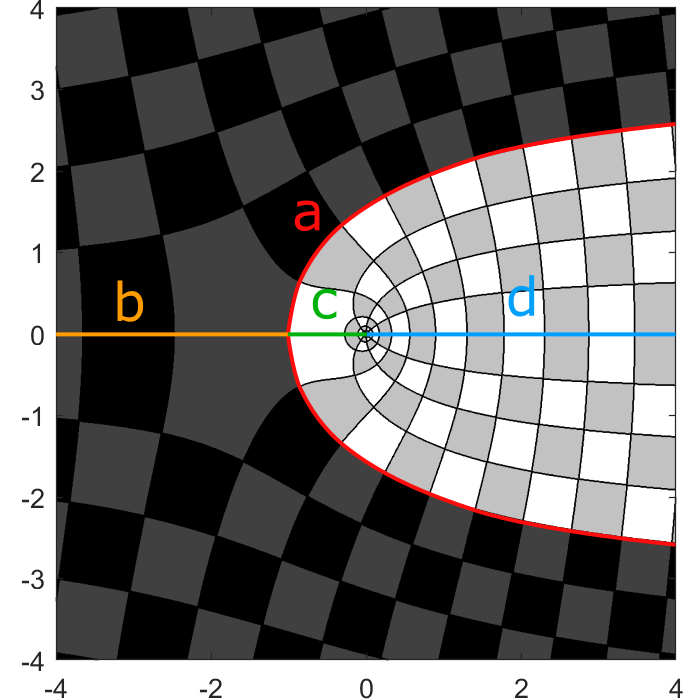}
		\caption{}
		\label{figure_grid_plume}
	\end{subfigure}
	\begin{subfigure}{0.45\textwidth}
		\includegraphics[width=\textwidth]{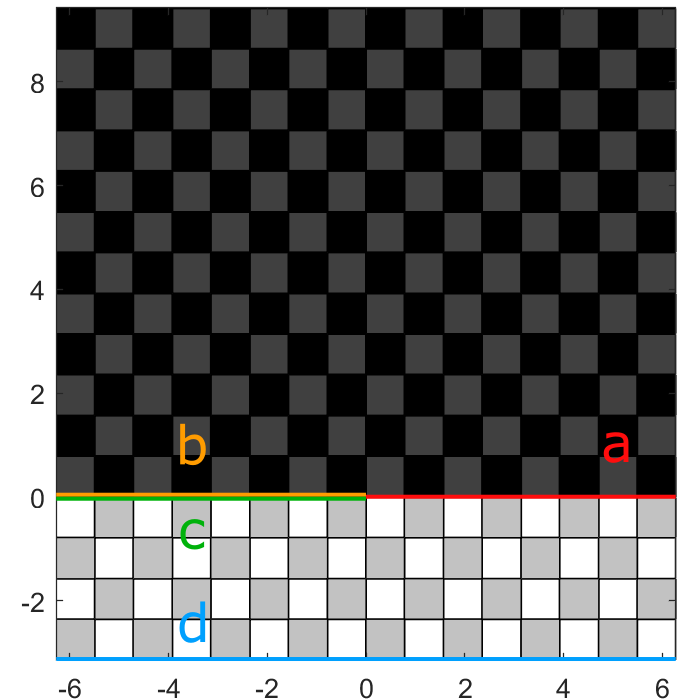}
		\caption{}
		\label{figure_flat_grid_plume}
	\end{subfigure}

	\caption{Map of the important features of the thermal plume domain (\subref{figure_grid_plume}) and their equivalent in the streamline coordinate domain (\subref{figure_flat_grid_plume})}
	\label{figure_streamline_plume}
\end{figure}

We can once again map the important segments of the problem to the streamline coordinate domain. This mapping is illustrated in figure \ref{figure_streamline_plume}. An explicit map of each segment is presented in table \ref{table_plume_features}. In both instances we use the map $\phi + i \psi = \Phi + \Phi_{stag}$ so that the image of the stagnation point lands at the origin of the streamline domain.

\begin{table}
  \begin{center}
  \begin{tabular}{cccc}
	\hline
	Segment 	&	Z-plane coordinates 	& Streamline coordinates 	& 	Boundary condition \\ \hline
 	a)			& $r = \frac{\pi - \alpha \theta}{\text{sin} \left( \theta \right)}$					& $\phi \ \in \ ] 0, \infty[$		& Continuity 	\\
				& 								& $\psi = 0$	& 	\\ \hline
	b)			& $x \ \in \ ] -\infty, 1 [$		& $\phi \ \in \ ] -\infty,0[$	& Symmetry 	\\
				& $y = 0$						& $\psi = 0^+$	&  	\\ \hline
	c)			& $x \ \in \ ]-1, 0[$				& $\phi \ \in \ ]-\infty, 0[$	& Symmetry 	\\
				& $y = 0$						& $\psi = 0^-$				&  	\\ \hline
	d)			& $x \ \in \ ]0, x_{stag}[$		& $\phi \ \in \ ]-\infty, \infty[$		& Symmetry 	\\
				& $y = 0$						& $\psi =- \alpha \pi$				&  	\\ \hline
  \end{tabular}
  \caption{Important features of the thermal plume domain and their equivalent in the streamline coordinate domain. Z-plane expression for segment a) is expressed in polar coordinates centered on the injection aperture for simplicity}
  \label{table_plume_features}
  \end{center}
\end{table}

The first high $\Pen$ approximation \ref{solution_wake} is applicable directly to this problem. To obtain a high $\Pen$ solution valid far downstream, we must modify the results of section \ref{section_high_Pe_2} to account for the fact that the domain is now semi-infinite in the $\psi$ direction. This solution can again be obtained by using the Green’s functions for Helmholtz’ equation in the quadrant, with Dirichlet boundary conditions on one axis and Neumann boundary conditions on the other. Because the domain in streamline coordinates is semi-infinite, the low $\Pen$ regime is harder to analyze than the strip domain of the dipole solution. We leave to future work the determination of the diffusion profile for the thermal plume at very low values of $\Pen$, which would have to be done using matched asymptotic expansions (as the approximation of complete mixing in the $\psi$ direction has to break down for sufficiently high $\psi$).

\begin{figure}
	\centering
	\begin{subfigure}{0.45\textwidth}
		\includegraphics[width=\textwidth]{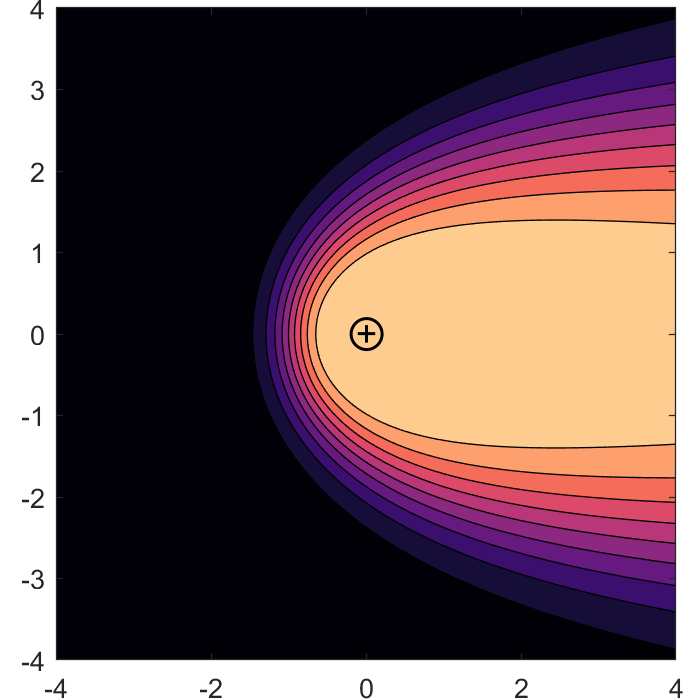}
		\caption{}
		\label{figure_plume_concentration}
	\end{subfigure}
	\begin{subfigure}{0.45\textwidth}
		\includegraphics[width=\textwidth]{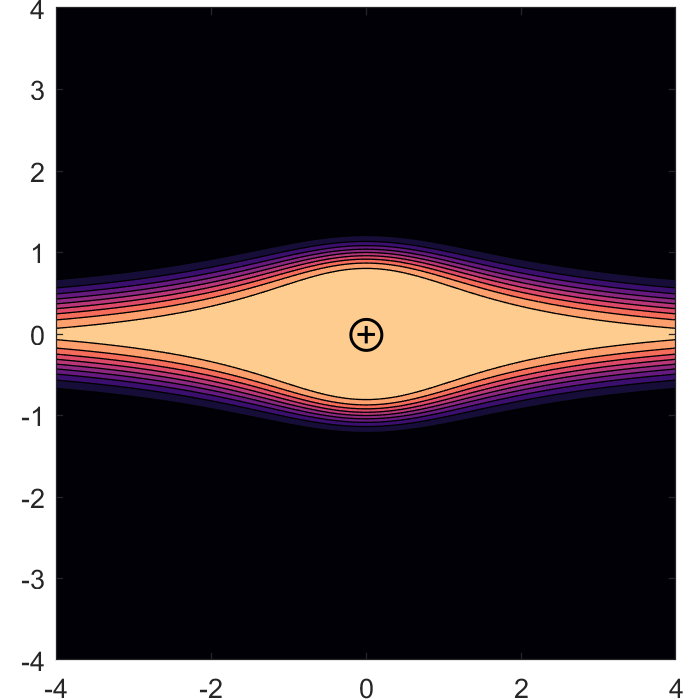}
		\caption{}
		\label{figure_plume_extensional}
	\end{subfigure}

	\caption{Concentration for a thermal plume in a straight flow with $\Pen = 10$ and $\alpha = 1$ (\subref{figure_grid_dipole}) and concentration for a single injection aperture superimposed over a purely extensional flow (\subref{figure_flat_grid_dipole}), obtained by applying the power transform $z = z^{'2}$ to the solution illustrated in (\subref{figure_plume_concentration})}
	\label{figure_streamline}
\end{figure}

An example of high $\Pen$ solution for a thermal plume in a straight flow is illustrated in figure \ref{figure_plume_concentration}. The solution can again be transformed using simple conformal maps to obtain concentration profiles for more elaborate flow patterns. This is illustrated in figure \ref{figure_plume_extensional}, where we use a simple power transform on the plume solution to obtain the concentration profile of a single injection aperture placed over a purely extensional flow.

\section{Concluding remarks}

We have analyzed the problem of tracer dispersion in 2D multipolar flows and shown solutions to the problem for both high and low $\Pen$ limits. We have shown two high $\Pen$ solutions, valid respectively downstream and upstream, and which can be combined to obtain an approximation valid everywhere in the flow domain. We have shown how simple solutions for transport in multipolar devices can be extended to obtain 2D concentration profile in a variety of geometries. These solutions allow for a more sophisticated, 2-dimensional description of convection-diffusion in many transport problems. Specifically, they give us complete concentration profiles in domains where we previously relied on material surface tracking or limited scaling arguments. These include open-space microfluidic applications such as microfluidic probes or multipoles, reinjection problems in groundwater flows found in groundwater heat pumps or aquifer contaminations and in interface tracking problems in porous media. Moreover, the technique demonstrated here is not limited to the few arrangements of sources and sinks we have illustrated. By using the same transformations with more elaborate initial geometries, we could generate concentration profiles for increasingly sophisticated devices. We hope that a more thorough knowledge of 2D exchange phenomena in open-space system will allow engineers and experimentalists from a wide range of fields to design devices that exploit the full features of 2D convection-diffusion, without being limited by unidirectional lump models.

\paragraph{Acknowledgements.}
E.B. acknowledges funding from the Fonds de Recherche du Qu\'ebec (FRQ) "Bourse de doctorat en recherche" and from the National Science and Engineering Research Council of Canada (NSERC) "Bourse Alexander Graham Bell". T.G. acknowledges funding from the Fonds de Recherche du Qu\'ebec (FRQ), "\'Etablissement de nouveaux chercheurs" and "\'Equipe" programs, and the National Science and Engineering Research Council of Canada (NSERC – RGPIN - 06409). We thank the Canadian Microsystems Center (cmc.ca) for access to a shared computational infrastructure.

\paragraph{Declaration of Interests.}
The authors report no conflict of interest.



\bibliographystyle{jfm}
\bibliography{multipole_reseparated}

\end{document}